\documentclass[conference]{IEEEtran}
\usepackage[noadjust]{cite}
\def\3To1BC{$3-$to$-1$}

\def\define{:{=}~}
\def\naturals{\mathbb{N}}

\def\underlineY{\underline{Y}}

\def\deltah{\partial h_{b}}
\def\cocl{\mbox{cocl}}

\def\SetOfDistributions{\mathbb{D}}

\def\TimeSharingRVSet{\mathcal{Q}}
\def\TimeSharingRV{Q}

\def\SemiPrivateRVSet{\mathcal{U}}

\def\InputRV{X}

\def\SemiPrivateRV{U}

\def\Expectation{\mathbb{E}}

\def\setX{\mathcal{X}}
\def\setS{\mathcal{S}}

\def\threeIC{$3-$IC }
\def\2IC{$2-$IC}

\def\cost{\tau}
\def\ulinecost{\underline{\cost}}

\def\three-1{3\mbox{-}1}

\def\\3to1IC{$3-$to$-1$ IC}
\def\threeto1{$3-$to$-1$}

\def\ulineInputAlphabet{\underline{\InputAlphabet}}
\def\ulineOutputAlphabet{\underline{\OutputAlphabet}}
\def\2IC{$2-$IC}
\def\OutputRV{Y}
\def\ulineOutputRV{\underline{\OutputRV}}

\def\ulineInputRV{\underline{\InputRV}}

\def\ulineoutput{\underline{y}}
\def\ulineinput{\underline{x}}

\def\costfn{\kappa}
\def\ulinecostfn{\underline{\costfn}}

\usepackage{amssymb}
\usepackage{amsmath}
\usepackage{mathrsfs}
\usepackage{ulem}
\usepackage{epsf,epsfig}

\newcommand{\InputAlphabet}{\mathcal{X}}
\newcommand{\OutputAlphabet}{\mathcal{Y}}

\newcommand{\comment}[1]{}
\begin{document}
\sloppy
\newtheorem{remark}{\it Remark}
\newtheorem{thm}{Theorem}
\newtheorem{corollary}{Corollary}
\newtheorem{definition}{Definition}
\newtheorem{lemma}{Lemma}
\newtheorem{example}{Example}
\newtheorem{prop}{Proposition}

\title{Coset codes for communicating over non-additive channels}

\author {Arun~Padakandla and S. Sandeep~Pradhan~\IEEEmembership{Member,~IEEE}
\thanks{Arun Padakandla and S. Sandeep Pradhan are with the Department of Electrical
and Computer Engineering, University of Michigan, Ann Arbor
48109-2122, USA.}
\thanks{This work was supported by NSF grant CCF-1116021.}}

\maketitle
\begin{abstract}
We present a case for the use of codes possessing algebraic closure properties - coset codes - in developing coding techniques and characterizing achievable rate regions for \textit{generic} multi-terminal channels. In particular, we consider three diverse communication scenarios - $3-$user interference channel (many-to-many), $3-$user broadcast channel (one-to-many), and multiple access with distributed states (many-to-one) - and identify \textit{non-additive} examples for which coset codes are \textit{analytically} proven to yield strictly larger achievable rate regions than those achievable using iid codes. On the one hand, our findings motivate the need for multi-terminal information theory to step beyond iid codes. On the other, it encourages current research of linear code-based techniques to go beyond particular additive communication channels. Detailed proofs of our results are available in \cite{201403arXiv_PadPra, 201207arXiv_PadPra, 201301arXivMACDSTx_PadPra}.
\end{abstract}
\section{Introduction and Preliminaries}
\label{Sec:Introduction}

Proving achievability of rate regions via random coding is synonymous with the use of iid codebooks. 
Successes in the context of point-to-point (PTP), multiple access (MAC) and particular multi-terminal channels such as degraded broadcast channels (BCs) have fueled a widely held belief that if computation were a no-issue, then one can achieve capacity using iid codebooks, or in other words, codebooks possessing simple single-letter empirical properties. Brought to light over three decades ago, K\"orner and Marton's \cite{197903TIT_KorMar} technique based on statistically dependent codebooks possessing algebraic closure properties, henceforth referred to as structured codebooks, outperformed all current known techniques based on iid codebooks and challenged this widely held belief. More recently, similar findings \cite{200710TIT_NazGas, 200809Allerton_SriJafVisJafSha, 200906TIT_PhiZam, 201308TIT_NieMad, 201407TIT_KriJaf, 201408TIT_HonCai} have reinforced the utility of algebraic closure properties in the context of particular \textit{symmetric and additive} multi-terminal communication scenarios. Though these findings present an encouraging sign and a new tool to attack long standing multi-terminal information theory problems, the use of structured codes remains outside mainstream information theory and is met with skepticism.

Among others, three primary reasons for this are the following. Firstly, in contrast to the rich theory based on iid codebooks, structured codes have been studied only in the context of particular additive and symmetric channels.\footnote{An exception to this is \cite{201103TIT_KriPra} wherein K\"orner and Marton's technique is generalized to an arbitrary distributed source coding problem.} In other words, the lack of a general theory - an achievable rate region based on structured codes for arbitrary instances of the multi-terminal channel in question - fuels doubt. Secondly, the lack of a rich set of examples, beyond particular symmetric additive examples for which structured codes outperform current known techniques based on iid codes increases skepticism. Lastly, the lack of wider applicability of structured codes to diverse communication scenarios, for ex. BCs - a one-to-many communication scenario - also adds to doubt.\footnote{Indeed, benefits of structured codes are known only for many-to-one communication scenarios and certain function computation problems.}

In this article, we lay to rest the above doubts by presenting non-additive examples for which structured codes strictly outperform iid codebooks. In particular, we present \textit{non-additive} examples for three diverse communication scenarios - $3-$user interference channel ($3-$IC), $3-$user broadcast channel ($3-$BC) and a MAC with channel state information distributed at transmitters (MAC-DSTx) - and \textit{analytically} prove structured code based techniques yield strictly larger achievable rate regions than those based on iid codebooks.

In section \ref{Sec:3To1ICs}, we build on \cite{201403arXiv_PadPra} to indicate how alignment \cite{200809Allerton_SriJafVisJafSha, 201009TIT_BreParTse, 201407TIT_KriJaf, 201408TIT_HonCai} can be performed, and is beneficial, for non-additive $3-$ICs.\footnote{We remark that current alignment techniques are restricted to \textit{additive} $3-$ICs.} Of particular interest is Ex. \ref{Ex:3-to-3}, wherein we demonstrate that our technique can effect alignment at all receivers simultaneously, even when the underlying alphabet set is finite. The use of structured codes for BCs was initiated in \cite{201207arXiv_PadPra}, wherein the first example for which coding techniques based on \cite{197905TIT_Mar} were proven to be sub-optimal. Going beyond this additive example, we present a non-additive $3-$BC in section \ref{Sec:3UserBC} for which structured codes are strictly more efficient.

Providing analytical proofs for strict containment of iid code based techniques is fraught with challenges. For `non-standard' instances, such as the non-additive ones considered here, there are no techniques for evaluating achievable rate regions without resorting to computation. Owing to loose bounds on auxiliary alphabet sets, the latter is not feasible with current computation power. In fact, even in the case of additive examples, strict sub-optimality of iid code based techniques are proven only in a handful of communication scenarios. In our work, we devise a new line of argument to overcome these challenges without resorting to computation.

The significance of our work is summarized as follows. First and foremost, through our examples, we provide a definitive reasoning to step beyond iid codebooks and adopt ensembles of codes possessing richer properties. Given that achievability proofs are synonymous with iid codebooks, the import of this cannot be overstated. Secondly, our non-additive examples validate the need to go beyond our current understanding of structured codes for particular additive and symmetric instances and develop a theory for generic multi-terminal channels. Thirdly, the analytical techniques we develop to prove strict sub-optimality of iid code based techniques might be useful for similar endeavors in other settings.


We employ notation that is standard in information theory literature supplemented by the following. For $K \in \naturals$, we let $[K]\define \left\{ 1,2\cdots,K \right\}$. We let $BSC_{\eta}(0|1)=BSC_{\eta}(1|0)=1-BSC_{\eta}(0|0)=1-BSC_{\eta}(1|1)=\eta$ denote the transition probabilities of a BSC. We let $h_{b}(x) = -x\log_{2}x-(1-x)\log_{2}(1-x)$ denote binary entropy function, $a*b = a(1-b)+(1-a)b$ denote binary convolution, calligraphic letters such as $\mathcal{X}, \mathcal{Y}$ denote finite sets. Let $\mathcal{F}_{q}$ denote a finite field of cardinality $q$ and $\oplus_{q}$ addition in $\mathcal{F}_{q}$. Let $\deltah(\tau,\delta)\define h_{b}(\tau*\delta)-h_{b}(\delta)$ denote capacity of a binary symmetric channel (BSC) with cross over probability $\delta$ and Hamming cost constraint $\tau$. We use an \underline{underline} to denote aggregates of objects of similar type. For example, a $3-$IC with input alphabets $\mathcal{X}_{j}:j=1,2,3$, output alphabets $\mathcal{Y}_{j}:j=1,2,3$, channel transition probabilities $W_{Y_{1}Y_{2}Y_{3}|X_{1}X_{2}X_{3}}$ and channel input cost functions $\kappa_{j}: \mathcal{X}_{j} \rightarrow \mathbb{R}:j=1,2,3$ is referred to as the $3-$IC $(\ulineInputAlphabet,\ulineOutputAlphabet,W_{\ulineOutputRV|\ulineInputRV},\ulinecostfn)$.

\section{$3-$user interference channels}
\label{Sec:3To1ICs}
All the $3-$IC's studied in this article are binary $3-$ICs with Hamming cost functions, i.e., $\mathcal{X}_{j}=\mathcal{Y}_{j}= \{0,1\}$ $\kappa_{j}=\kappa_{H}$, where $\kappa_{H}(x)=x$ for all $x \in \mathcal{X}_{j}= \{0,1\}$ for all $j=1,2,3$.
\subsection{$3-$to$-1$ Interference Channels}
\label{SubSec:3To1IC}
We begin with examples of $3-$to$-1$ ICs - a collection of $3-$IC's wherein only one of the users suffers from interference, and the other two users enjoy interference-free PTP channels \cite{201009TIT_BreParTse, 200912arXiv_CadJaf}. Since interference is isolated to a single receiver (Rx) in a $3-$to$-1$ IC, it lets us highlight the drawbacks of current known techniques based on iid codes for interference mitigation.

\begin{figure}
\begin{minipage}{.5\textwidth}
\centering\includegraphics[height=1.7in]{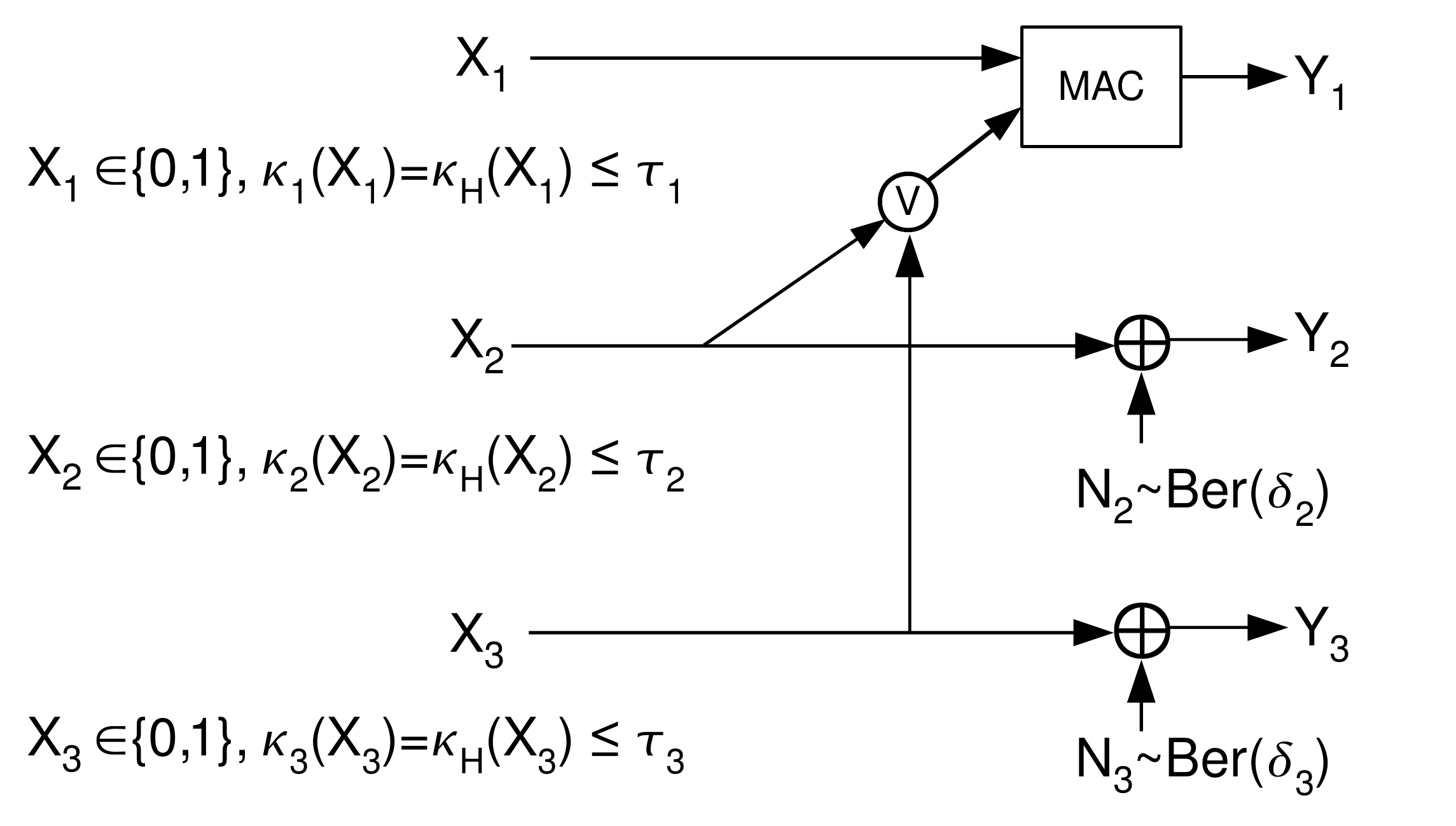}
\caption{A binary $3-$to$-1$ IC described in examples \ref{Ex:A3To1-OR-IC} and \ref{Ex:3To1ORICCoupledThroughNonAdditiveMAC}.}
\label{Fig:3To1ORICWithNonAdditiveMAC}
\end{minipage}
\end{figure}

\begin{example}
\label{Ex:A3To1-OR-IC}
Consider a binary \threeIC illustrated in figure \ref{Fig:3To1ORICWithNonAdditiveMAC} wherein the MAC depicted is a binary additive MAC with cross over probability $\delta_{1}$. Formally, $W_{\ulineOutputRV|\ulineInputRV}(\ulineoutput|\ulineinput)=BSC_{\delta_{1}}(y_{1}|x_{1}\oplus (x_{2} \vee x_{3}))BSC_{\delta_{2}}(y_{2}|x_{2})BSC_{\delta_{3}}(y_{3}|x_{3})$, where $\vee$ denotes logical OR. User $j$th input is constrained to an average Hamming cost $\tau_{j} \in (0,\frac{1}{2})$ per symbol for $j \in [3]$.
\end{example}

The interference - $X_{2}\vee X_{3}$ - seen by Rx $1$ is a non-additive function of $X_{2},X_{3}$. Since logical OR and binary addition are the only two non-trivial bivariate binary functions, they can be viewed to be at two ends of a spectrum. Therefore the former is `as non-additive a function as it can get'. In the following, we argue coset codes built over finite fields strictly outperform iid codes even for this non-additive $3-$to$-1$ IC. Ex. \ref{Ex:A3To1-OR-IC} is studied in \cite[Ex. 2]{201403arXiv_PadPra}, and a detailed proof of the above statement is provided in \cite[Appendix G]{201403arXiv_PadPra}.

Our strategy to establish the above, relies on the structure of Ex. \ref{Ex:A3To1-OR-IC}. We derive conditions under which (i) iid codes do \textit{not} permit each of the Rxs to achieve their PTP capacities simultaneously and (ii) coset codes permit the same. We then identify an instance of Ex. \ref{Ex:A3To1-OR-IC} that satisfy these conditions by explicitly assigning values for $\delta,\tau,\delta_{1},\tau_{1}$. Let us begin by investigating how current known techniques based on iid codes attempt to achieve PTP capacity simultaneously for each user.

Since users $2,3$'s transmissions cause interference to Rx $1$, they split their transmission into two parts via superposition coding \cite{197303TIT_Ber}. For $j=2,3$, let $U_{j}$ and $X_{j}$ denote cloud center and satellite codebooks respectively. Since user $1$ does not cause interference to any Rx, it does not split it's transmission $X_{1}$. Rx $1$ decodes $U_{2},U_{3}, X_{1}$ and Rxs $2,3$ decode $U_{2},X_{2}$ and $U_{3},X_{3}$ respectively. It can be verified that the maximum rate achievable by user $1$ is $I(X_{1};Y_{1}|U_{2},U_{3})$. Given that $Y_{1}= X_{1} \oplus (X_{2}\vee X_{3})\oplus N_{1}$, where $N_{1}$ is a Bernoulli($\delta_{1}$) noise process, and $X_{1}$ is Hamming cost constrained to $\tau_{1}$, it can be verified that the upper bound $I(X_{1};Y_{1}|U_{2},U_{3})$ is strictly lesser than $\deltah(\tau_{1},\delta_{1})$, the PTP capacity of user $1$, unless $H(X_{2} \vee X_{3}|U_{2},U_{3}) = 0$ and $p_{X_{1}}(1) = \tau_{1}$.

When can $H(X_{2} \vee X_{3}|U_{2},U_{3}) = 0$? In order to achieve PTP capacities of users $2$ and $3$, $X_{2}$ and $X_{3}$ must be non-degenerate and independent.\footnote{In fact, we need $p_{X_{2}}(1) = p_{X_{3}}(1) = \tau \in (0,\frac{1}{2})$.} In this case, $H(X_{2}\vee X_{3}|U_{2},U_{3}) = 0$ iff $H(X_{j}|U_{j}) = 0$ for $j=2,3$. The latter condition implies Rx $1$ must decode entire transmissions of users $2,3$. This is possible only if the rates of the three users $R_{1},R_{2},R_{3}$ satisfy $R_{1}\!+\!R_{2}\!+\!R_{3}\! < \!I(X_{1}X_{2}X_{3};Y_{1})$. Substituting for distributions of $X_{1},X_{2},X_{3}$ that are necessary for achieving their PTP capacities and $R_{j} = \deltah(\tau_{j},\delta_{j})$ for $j \in [3]$ in the above inequality, we obtain a necessary condition for the above technique to be able to achieve PTP capacities for each user simultaneously.
\begin{prop}
 \label{Prop:ORICStrictSub-OptimalityOfHK}
Consider the \\3to1IC described in Ex. \ref{Ex:A3To1-OR-IC} with $\delta \define \delta_{2}=\delta_{3} \in (0,\frac{1}{2})$ and $\tau \define \tau_{2}=\tau_{3} \in (0,\frac{1}{2})$. Let $\beta \define \delta_{1}*(2\tau -\tau^{2})$. If
\begin{equation}
 \label{Eqn:3To1-OR-ICConditionForStrictSubOptimalityOfHKInProp}
\deltah(\tau_{1},\delta_{1})+2(\deltah(\tau,\delta))> h_{b}(\tau_{1}*\beta)-h_{b}(\delta_{1}),
\end{equation}
then the rate triple $(\deltah(\tau_{1},\delta_{1}),\deltah(\tau ,\delta),\deltah(\tau,\delta))$ is \textit{not} achievable using iid codes.
\end{prop}

Though the interference is a bivariate function - $X_{2}\vee X_{3}$ - of $X_{2},X_{3}$, iid codes force user $1$ to infer the interference by decoding separate univariate components $U_{2},U_{3}$. In our article \cite{201403arXiv_PadPra}, we propose Rx $1$ decode bivariate functions of cloud center codebooks $U_{2},U_{3}$. Specifically, consider the above coding technique with cloud center codebooks $U_{2},U_{3}$ being cosets of a linear code built over a common finite field $\mathcal{U}_{2} = \mathcal{U}_{3} = \mathcal{F}_{q} $. As before, Rxs $2$ and $3$ decode $U_{2},X_{2}$ and $U_{3},X_{3}$ respectively. Rx $1$ decodes $U_{2}\oplus_{q} U_{3},X_{1}$. The joint structure of cloud center codebooks restricts the number of $U_{2}\oplus_{q}U_{3}$ sequences and thereby efficient decoding of the same. This coding technique yields an achievable rate region that is characterized in \cite[Thm 2]{201403arXiv_PadPra}. In here, we only discuss how decoding a linear function of cloud center codebooks enables Rx $1$ to efficiently infer the non-linear interference in Ex. \ref{Ex:A3To1-OR-IC}.

The key fact is that though $X_{2}\vee X_{3}$ is non-linear over the binary field, it can be inferred from a linear function over a larger finite field. For example, pretend that $X_{2},X_{3}$ take values over the ternary field $\mathcal{F}_{3}$ (with $P(X_{2}=2)=P(X_{3}=2)=0$). Since $H(X_{2}\vee X_{3}|X_{2}\oplus_{3}X_{3}) = 0$, user $1$ can reconstruct the interference by decoding the ternary sum $X_{2}\oplus_{3}X_{3}$. This indicates that if we were to choose $U_{2},U_{3} \in \mathcal{F}_{3}$ with $U_{j}=X_{j}$, $1-P(U_{j}=0) =P(U_{j}=1)=\tau$ for $j=2,3$ and $P_{X_{1}}(1)=\tau_{1}$, then (i) $H(X_{2}\vee X_{3}|U_{2}\oplus_{3}U_{3}) = 0$ and (ii) $X_{j}:j \in [3]$ possess capacity achieving distributions, and therefore the above coding technique supports PTP capacity for each user simultaneously.\footnote{In the interest of brevity, we have glossed over details such as achieving capacity of PTP channels of users $2,3 $ using coset codes, etc. We refer the reader to \cite[Example 2]{201403arXiv_PadPra}, where all of these elaborated upon.} In the following proposition, we state condition on parameters for this to hold.

\begin{prop}
 \label{Prop:ORICStrictSub-OptimalityOfHK}
Consider the \\3to1IC described in example \ref{Ex:A3To1-OR-IC} with $\delta \define \delta_{2}=\delta_{3} \in (0,\frac{1}{2})$ and $\tau \define \tau_{2}=\tau_{3} \in (0,\frac{1}{2})$. Let $\beta \define \delta_{1}*(2\tau -\tau^{2})$. If
\begin{eqnarray}
 \label{Eqn:3To1-OR-ICConditionForAchievabilityInProp}
\deltah(\tau,\delta)\leq  \theta ,
\end{eqnarray}
where $\theta =h_{b}(\tau)-h_{b}((1-\tau)^{2})-(2\tau-\tau^{2})h_{b}(\frac{\tau^{2}}{2\tau-\tau^{2}})-h_{b}(\tau_{1} * \delta_{1})+h_{b}(\tau_{1}*\beta)$, then $(\deltah(\tau_{1},\delta_{1}),\deltah(\tau ,\delta),\deltah(\tau,\delta))$ is achievable using coset codes.
\end{prop}

Conditions (\ref{Eqn:3To1-OR-ICConditionForAchievabilityInProp}) and (\ref{Eqn:3To1-OR-ICConditionForStrictSubOptimalityOfHKInProp}) are \textit{not} mutually exclusive. It maybe verified that the choice $\tau_{1}=\frac{1}{90}$, $\tau=0.15$, $\delta_{1}=0.01$ and $\delta=0.067$ satisfies both conditions, thereby establishing the utility of structured codes for non-additive $3-$to$-1$ IC of example \ref{Ex:A3To1-OR-IC}.

Our goal now is to go one more step and replace the binary additive MAC in example \ref{Ex:A3To1-OR-IC} with a non-additive one. 

\begin{example}
 \label{Ex:3To1ORICCoupledThroughNonAdditiveMAC}
Consider a binary \\3to1IC depicted in figure
\ref{Fig:3To1ORICWithNonAdditiveMAC} with channel transition probabilities
$W_{\ulineOutputRV|\ulineInputRV}(\ulineoutput|\ulineinput)=MAC(y_{1}|x_{1},
x_{2} \vee x_{3})BSC_{\delta}(y_{2}|x_{2})BSC_{\delta}(y_{3}|x_{3})$,
where $MAC(0|0,0)=0.989, MAC(0|0,1)=0.01, MAC(0|1,0)=0.02,
MAC(0|1,1)=0.993$ and $MAC(0|b,c)+MAC(1|b,c)=1$ for each $(b,c) \in \{
0,1 \}^{2}$. User $j$th input is constrained to an average Hamming cost $\tau_{j} \in (0,\frac{1}{2})$ per symbol, where $\tau\define\tau_{2}=\tau_{3}$.
\end{example}

How does one \textit{analytically} prove strict sub-optimality of iid codes for the above example? The reader will recognize that the MAC being `non-standard', this is significantly harder. Our proof closely follows the line of argument presented for example \ref{Ex:A3To1-OR-IC}, thereby validating the power of the technique presented therein.\footnote{The structure of a $3-$to$-1$ IC that captures the essential aspects in a simplified setting must not be overlooked.} Example \ref{Ex:3To1ORICCoupledThroughNonAdditiveMAC} is studied in \cite[Example 3]{201403arXiv_PadPra}, and a detailed proof (of proposition \ref{Prop:ResultForOrExampleWithMACAsAProp}) is provided in \cite[Appendix H]{201403arXiv_PadPra}. In the following, we only highlight how the argument for example \ref{Ex:3To1ORICCoupledThroughNonAdditiveMAC} differs from that of example \ref{Ex:A3To1-OR-IC}.

Observe that, the maximum rate achievable by user $1$ under a Hamming constraint of $\tau_{1}$, given that users $2,3$ achieve their PTP capacities, is
\begin{eqnarray}
 \label{Eqn:TestChannelsThatEnsureUsers2And3AchieveCapacity}
\lefteqn{C_{1} \define \underset{p_{\underline{X}\ulineOutputRV} \in \mathcal{D}(\ulinecost )}{\sup} I(X_{1};Y_{1}|X_{2}\vee X_{3}),\mbox{ where,}}\\ 
\!\!\!\!\!\!\!\!\label{Eqn:TestChannelsNonAdditive3To1IC}&\!\!\!\!\!\!\!\!\mathcal{D}(\ulinecost) \define \left\{ \!\!\!\begin{array}{c}p_{\ulineInputRV\ulineOutputRV} \mbox{ is a pmf on }\ulineInputAlphabet\times\ulineOutputAlphabet : p_{\ulineOutputRV|\ulineInputRV}=W_{\ulineOutputRV|\ulineInputRV},\\p_{\ulineInputRV}=p_{X_{1}}p_{X_{2}}p_{X_{3}}, p_{X_{j}}(1)=\tau\mbox{ for }\\j=2,3 \mbox{ and }p_{X_{1}}(1)\leq \tau_{1}\end{array}\!\!\!\right\}.
\end{eqnarray}
$C_{1}$, and $p^{*}_{\ulineInputRV\ulineOutputRV} \in \mathcal{D}(\ulinecost)$ that achieves $C_{1}$, can be numerically computed in quick time. A careful reader will now recognize that we can essentially retrace our arguments for Ex. \ref{Ex:A3To1-OR-IC} by substituting $C_{1}$ and $p^{*}_{\ulineInputRV\ulineOutputRV}$ for $\deltah(\tau_{1},\delta_{1})$ and the capacity achieving distribution therein. Specifically, we can derive conditions under which the rate triple $\underline{C}^{*}\define (C_{1},\deltah(\tau ,\delta),\deltah(\tau,\delta))$ is (i) \textit{not} achievable using iid codes, \textit{and} (ii) is achievable using coset codes. We then show that these conditions can be satisfied by an explicit assignment for $\delta,\tau_{1},\tau$.

\begin{prop}
 \label{Prop:ResultForOrExampleWithMACAsAProp}
Consider example \ref{Ex:3To1ORICCoupledThroughNonAdditiveMAC} and let $\underline{C}^{*}, C_{1}, \mathcal{D}(\ulinecost),p^{*}_{\ulineInputRV\ulineOutputRV}$ be defined as above. If
\begin{equation}
 \label{Eqn:3To1OrNonAddICUnstructuredCodesSubOptimalInProp}
I(\ulineInputRV ;Y_{1})\! <\! I(X_{1};Y_{1}|X_{2}\!\vee\! X_{3})+2\deltah(\tau,\delta)\!=\!C_{1}\!+\!2\deltah(\tau,\delta)\nonumber
\end{equation}
where $I(X_{1};Y_{1}|X_{2}\vee X_{3})$, and $I(\ulineInputRV ;Y_{1})$ are evaluated with respect to $p^{*}_{\ulineInputRV\ulineOutputRV}$, then $\underline{C}^{*}$ is \textit{not} achievable using iid codes.
If
$h_{b}(\tau^{2})+(1-\tau^{2})h_{b}(\frac{(1-\tau)^{2}}{1-\tau^{2}})+H(Y_{1}|X_{2}\vee
X_{3})-H(Y_{1}) \leq \min\{ H(X_{2}|Y_{2}),H(X_{3}|Y_{3})\}$, where entropies are evaluated with respect to $p^{*}_{\ulineInputRV\ulineOutputRV}$, then $\underline{C}^*$ is achievable using coset codes.
\end{prop}

Please refer to \cite[Appendix H]{201403arXiv_PadPra} for a detailed proof. For example \ref{Ex:3To1ORICCoupledThroughNonAdditiveMAC}, with $\tau_{1}=0.01,\tau=\tau_{2}=\tau_{3}=0.1525,\delta = 0.067$, the conditions stated in proposition \ref{Prop:ResultForOrExampleWithMACAsAProp} hold simultaneously. For this channel, $p^{*}_{X_{1}}(0) = 0.99$,
\begin{eqnarray}
C_{1} +2(\deltah(\tau,\delta))- I(\ulineInputRV ;Y_{1}) = 0.0048, \mbox{ and}\nonumber\\
 h_{b}(\tau^{2})  +(1-\tau^{2}) h_{b}(\frac{(1-\tau)^{2}}{1-\tau^{2}}) + H(Y_{1}|X_{2}\vee X_{3})-H(Y_{1})
\nonumber\\
- \min\{ H(X_{2}|Y_{2})H(X_{3}|Y_{3})\} = - 0.0031 < 0.\nonumber
\end{eqnarray}

\subsection{$3-$user Interference Channels}
\label{SubSec:3UserIC}
\begin{figure}
\centering
\includegraphics[height=1.25in]{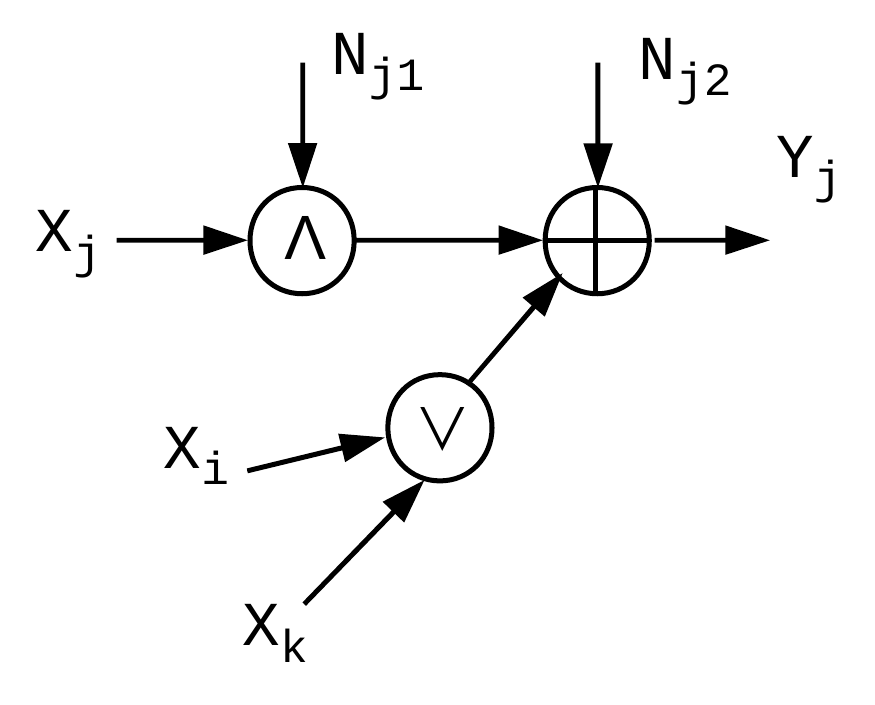}
\caption{The binary non-additive $3-$IC studied in example \ref{Ex:3-to-3}.}
\label{Fig:3-to-3}
\end{figure}

Is it possible to `align' interference over a generic $3-$IC wherein each user suffers from interference? Our next example indicates that this is indeed possible.\footnote{In general, aligning interference at multiple Rxs of a discrete $3-$IC is not possible and we conjecture a trade-off between the ability to communicate to one's own receiver and aid another by aligning \cite[Example 5]{201403arXiv_PadPra}.}
\begin{example}
\label{Ex:3-to-3}
Consider a binary $3-$IC whose inputs $X_{j}:j \in [3]$ and outputs $Y_{j}:j \in [3]$ are related as $Y_j=(X_j \land N_{j1}) \oplus (X_{i} \lor X_{k}) \oplus N_{j2}$ 
for $i,j,k \in[3]$, and $i,j$ and $k$ are distinct. This is depicted in figure \ref{Fig:3-to-3}.
$N_{ji}$, $j \in [3]$, $i \in [2]$ are mutually independent and independent of the inputs. 
$P(N_{j1}=1)=\beta$ and $P(N_{j2}=1)=\delta$ for $j \in [3]$. For $j \in [3]$, user $j$th input is constrained to an average Hamming cost $\tau$.
\end{example}

Let $i,j,k$ denote distinct indices in $[3]$. Can each user $j \in [3]$ achieve it's PTP capacity $I(X_{j};Y_{j}|X_{i}\vee X_{k})$ simultaneously? The coding technique based on coset codes described in the context of example \ref{Ex:A3To1-OR-IC} (prior to proposition \ref{Prop:CndsRteTrplAchvblUsingCosets}) can be generalized to $3-$IC by incorporating cloud center codebooks for each of the users and letting each receiver decode the sum of the other two cloud center codewords. We refer the reader to \cite[Thm 3]{201403arXiv_PadPra} for a characterization of the corresponding achievable rate region $\alpha_{f}(\underline{\tau})$. It can be verified that for the choice $\delta = 0.1, \tau = 0.1284,\beta = 0.2210$, each user can achieve it's PTP capacity simultaneously.

Note that, our findings for example \ref{Ex:3-to-3} (which have considerable practical significance) crucially relies on our generalization of alignment to arbitrary (including non-additive) $3-$ICs.

\section{$3-$user broadcast channel}
\label{Sec:3UserBC}
Let us paraphrase the key steps involved in proving proposition \ref{Prop:ORICStrictSub-OptimalityOfHK}. If Rx $1$ is unable to infer the interference $X_{2}\vee X_{3}$, then it cannot achieve it's PTP capacity. If Rx $1$ is constrained to decoding separate univariate components $U_{2},U_{3}$ of the user $2$ and $3$'s transmissions, then it cannot achieve it's PTP capacity unless $U_{2}=X_{2}$ and $U_{3}=X_{3}$. The channel parameters precludes receiver $1$ from decoding $X_{1},X_{2}=U_{2},X_{3}=U_{3}$ resulting in strict sub-optimality of iid code based techniques.

Can we bank on this argument to identify a (non-additive) $3-$BC for which iid codes are sub-optimal? Specifically, does the above argument hold for the $3-$BC obtained by pooling up the three inputs $X_{1},X_{2},X_{3}$ in example \ref{Ex:A3To1-OR-IC} as a single input $\ulineInputRV \define (X_{1},X_{2},X_{3})$ with three binary digits? We argue the answer is \textit{yes}, and we begin by stating the channel.

\begin{figure}
\centering
 \includegraphics[width=2.57in,height=1.5in]{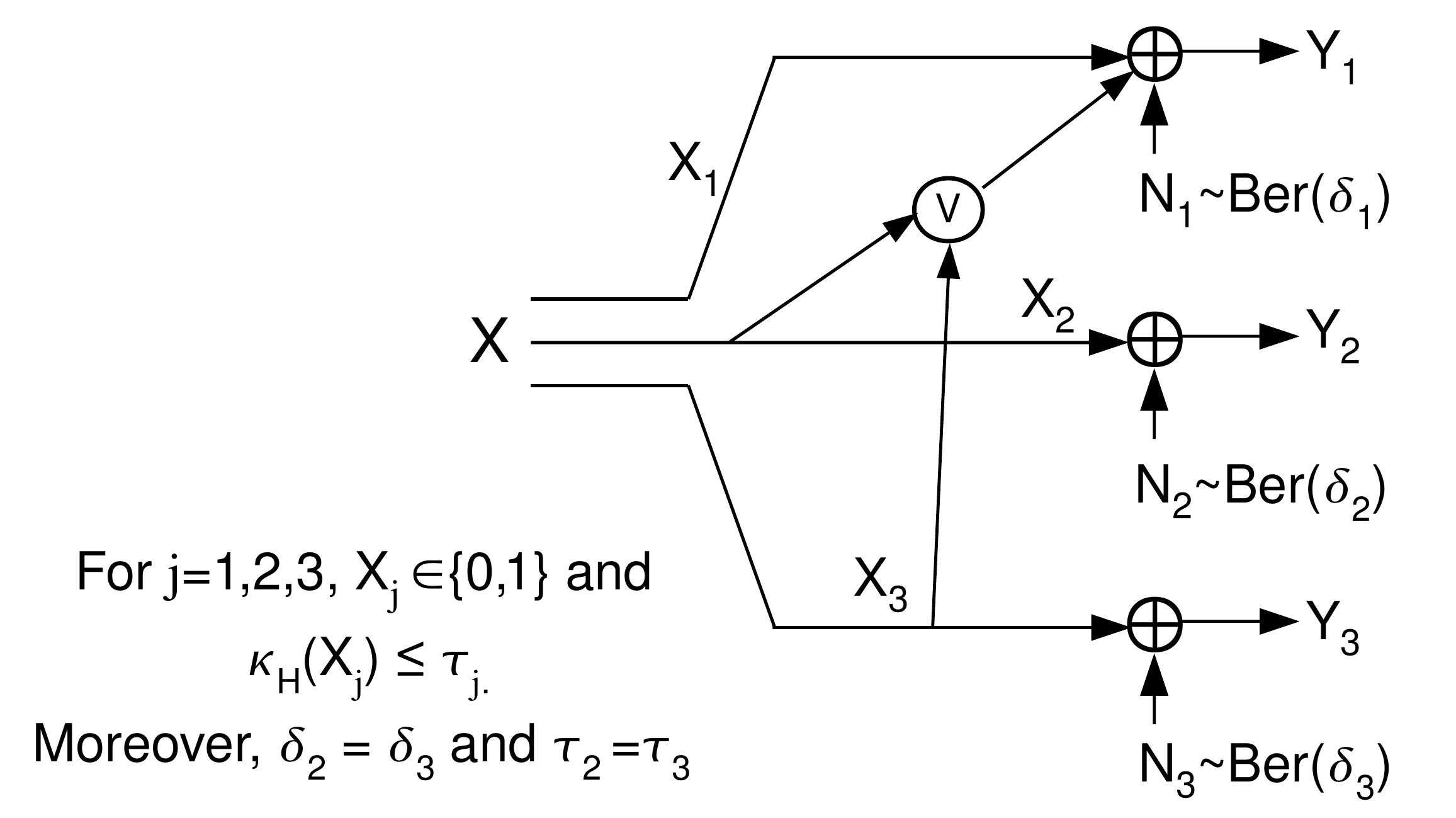}
\caption{The $3-$BC described in example \ref{Ex:A3To1-OR-BC}.}
\label{Fig:ORBC}
\end{figure}

\begin{example}
\label{Ex:A3To1-OR-BC}
Consider the $3-$BC depicted in fig. \ref{Fig:ORBC}, where the input alphabet $\InputAlphabet \define \left\{ 0,1 \right\}\times \left\{ 0,1 \right\} \times \left\{ 0,1 \right\}$, the output alphabets $\OutputAlphabet_{1}=\OutputAlphabet_{2}=\OutputAlphabet_{3} =\left\{ 0,1 \right\}$, and the channel transition probabilities $W_{\underlineY|X}(y_{1},y_{2},y_{3}|x_{1}x_{2}x_{3})=BSC_{\delta_{1}}(y_{1}|x_{1}\oplus (x_{2}\vee x_{3}))BSC_{\delta_{2}}(y_{2}|x_{2})BSC_{\delta_{3}}(y_{3}|x_{3})$ with $\delta \define \delta_{2}=\delta_{3}$. Each binary input digit is cost with respect to a Hamming cost function. Specifically, the cost function $\underline{\kappa} = (\kappa_{1},\kappa_{2},\kappa_{3})$, where $\kappa_{j} (x_{1}x_{2}x_{3}) = 1_{\left\{x_{j}=1\right\}}$ and the input $\ulineInputRV \define (X_{1},X_{2},X_{3})$ must satisfy $\Expectation\{\kappa_{j}(\ulineInputRV)\} \leq \tau_{j}$ for $j \in [3]$ with $\tau \define \tau_{2}=\tau_{3}$.
\end{example}
Please refer to \cite[Example 2]{201207arXiv_PadPra} for a study of Ex. \ref{Ex:A3To1-OR-BC} and proofs of propositions \ref{Prop:StrictSubOptimalityMartonFor3To1ORBC}, \ref{Prop:CndsRteTrplAchvblUsingCosets}. In here, we only describe the key ideas.

The binary digits $X_{2}$ and $X_{3}$ pass through interference free PTP channels to receivers $2$ and $3$ and these are the only digits through which receivers $2$ and $3$ can receive information. Suppose we require users $2,3$ to achieve their PTP capacities $\deltah(\tau,\delta)$, what is the maximum rate achievable by user $1$? It can be shown that the marginal distributions of $X_{2},X_{3}$ must be independent and satisfy $p_{X_{2}}(1)=p_{X_{3}}(1)=\tau$. Note that, unless the transmitter (Tx) utilizes it's knowledge of user $2$ and $3$'s codewords in choosing user $1$'s input $X_{1}$, it cannot communicate to user $1$ at it's PTP capacity $\deltah(\tau_{1},\delta_{1})$. This is because (i) the channel seen by receiver $1$ herein is identical to the channel seen by receiver $1$ in example \ref{Ex:A3To1-OR-IC}, and (ii) without using it's knowledge of user $2$ and $3$'s codewords, Tx $1$ is forced to live with superposition coding and the argument stated in the context of example \ref{Ex:A3To1-OR-IC} holds. 

However, using the knowledge of user $2$ and $3$'s codewords, Tx $1$ can precode \cite{1980MMPCT_GelPin, 197905TIT_Mar} for interference $X_{2}\vee X_{3}$. What is the maximum rate achievable via superposition coding and precoding\footnote{Superposition coding enables the Tx employ cloud center codebooks $U_{2},U_{3}$ for users $2,3$'s transmissions. The rest of the uncertainty $H(X_{2}\vee X_{3}|U_{2},U_{3})$ is precoded for.}? The reader will note that the equivalent channel seen by Rx $1$ is an additive PTP channel with channel state information \cite{1980MMPCT_GelPin} whose input $X_{1}$, channel state $S_{1}$ and output $Y_{1}$ are related as $Y_{1} = X_{1}\oplus S_{1} \oplus N_{1}$, where $N_{1}$ is a Bernoulli noise process with parameter $\delta_{1}$ and $S_{1}$ represents the residual uncertainty in the interference $X_{2}\vee X_{3}$ at Rx $1$ after it has decoded the cloud center codebooks $U_{2},U_{3}$. The key notion of \textit{rate loss} \cite{200305TIT_PraChoRam} implies that so long as $S_{1}$ is non-trivial and $X_{1}$ is constrained to a Hamming cost of $\tau_{1} \in (0,\frac{1}{2})$, user $1$ cannot achieve it's PTP capacity $\deltah(\tau_{1},\delta_{1})$. In other words, precoding does not let user $1$ achieve it's PTP capacity $\deltah(\tau_{1},\delta_{1})$ without perfect knowledge of interference $X_{2}\vee X_{3}$ at Rx $1$. We may now use the argument stated in the context of Ex. \ref{Ex:A3To1-OR-IC}, (paraphrased at the beginning of this section) to identify conditions on $\delta_{1},\tau_{1},\delta,\tau$ that preclude iid code based techniques from achieving PTP capacities for each user simultaneously. Not surprisingly, these conditions, stated in proposition \ref{Prop:StrictSubOptimalityMartonFor3To1ORBC}, are identical to those identified for Ex. \ref{Ex:A3To1-OR-IC}. Denoting $\alpha_{\mathscr{U}}(\ulinecost)$ as the current known largest achievable rate region for a $3-$BC using iid codes, we have the following.

\begin{prop}
 \label{Prop:StrictSubOptimalityMartonFor3To1ORBC}
Consider example \ref{Ex:A3To1-OR-BC} with $\delta \define \delta_{2}=\delta_{3} \in (0,\frac{1}{2})$ and $\tau \define \tau_{2}=\tau_{3} \in (0,\frac{1}{2})$. Let $\beta \define \delta_{1}*(2\tau -\tau^{2})$. The rate triple $(\deltah(\tau_{1},\delta_{1}),\deltah(\tau,\delta),\deltah(\tau,\delta)) \notin \alpha_{\mathscr{U}}(\underline{\tau})$ if
\begin{equation}
 \label{Eqn:3To1-OR-ICConditionForStrictSubOptimalityOfMarton}
\deltah(\tau_{1},\delta_{1})+2(\deltah(\tau,\delta))> h_{b}(\tau_{1}*\beta)-h_{b}(\delta_{1}).
\end{equation}
\end{prop}

A technique, similar in spirit to \cite{201403arXiv_PadPra}, is proposed in \cite{201207arXiv_PadPra}, wherein receiver $1$ decodes sum $U_{2}\oplus_{q}U_{3}$ of cloud center codebooks $U_{2},U_{3}$ taking values over finite field $\mathcal{F}_{q}$. By choosing cloud center codebooks to be cosets of a common linear code, the number of $U_{2}\oplus_{q}U_{3}$ sequences is squeezed, resulting in efficient decoding of the same. We denote the corresponding achievable rate region as $\beta_{1}(\ulinecost)$ whose characterization is provided in \cite[Defn 5, Thm 4]{201207arXiv_PadPra}. For the case of example \ref{Ex:A3To1-OR-BC}, we rely on a test channel analogous to the one employed for Ex. \ref{Ex:A3To1-OR-IC}. In particular, we let $U_{2},U_{3}$ live over the ternary field $\mathcal{F}_{3}$ and have $U_{2}=X_{2}$ and $U_{3}=X_{3}$ with probability $1$. Following earlier arguments, Rx $1$ can achieve it's PTP capacity $\deltah(\tau_{1},\delta_{1})$ if it can decode $U_{2}\oplus_{3}U_{3},X_{1}$. In the following proposition, we state conditions under which coset codes enable each user achieve it's PTP capacity.

\begin{prop}
 \label{Prop:CndsRteTrplAchvblUsingCosets}
Consider example \ref{Ex:A3To1-OR-BC} with $\delta \define
\delta_{2}=\delta_{3} \in (0,\frac{1}{2})$ and $\tau \define
\tau_{2}=\tau_{3} \in (0,\frac{1}{2})$. Let $\beta \define
\delta_{1}*(2\tau -\tau^{2})$. 
The rate triple $(\deltah(\tau_{1},\delta_{1}),\deltah(\tau,\delta),\deltah(\tau.\delta)) \in
\beta_{1}(\underline{\tau})$
i.e., achievable using coset codes, if, 
\begin{eqnarray}
 \label{Eqn:3To1-OR-BCConditionForAchievabilityUsingCosetCodes}
\deltah(\tau,\delta)\leq  \theta ,
\end{eqnarray}
where $\theta
=h_{b}(\tau)-h_{b}((1-\tau)^{2})-(2\tau-\tau^{2})h_{b}(\frac{\tau^{2}}{2\tau-\tau^{2}})-h_{b}(\tau_{1}
* \delta_{1})+h_{b}(\tau_{1}*\beta)$.
\end{prop}
Not surprisingly, we note that conditions (\ref{Eqn:3To1-OR-ICConditionForStrictSubOptimalityOfMarton}), (\ref{Eqn:3To1-OR-BCConditionForAchievabilityUsingCosetCodes}) are identical to conditions (\ref{Eqn:3To1-OR-ICConditionForStrictSubOptimalityOfHKInProp}), (\ref{Eqn:3To1-OR-ICConditionForAchievabilityInProp}). Therefore, the earlier choice $\tau_{1}=\frac{1}{90}, \tau=0.15, \delta_{1}=0.01, \delta=0.067$ satisfies both these conditions thereby establishing the utility of coset codes for non-additive $3-$BCs.
\section{Communicating over a MAC-DSTx}
\label{Sec:MACDSTx}

Consider a MAC analogue of a PTP channel with channel state (PTP-STx) studied by Gelfand and Pinsker \cite{1980MMPCT_GelPin}. For $j=1,2$, let $X_{j} \in \mathcal{X}_{j}$ denote encoder $j$'s input and $Y \in \mathcal{Y}$ denote the output. The channel transition probabilities depend on a random parameter $\underline{S} \define (S_{1},S_{2}) \in \underline{\mathcal{S}}\define \mathcal{S}_{1}\times \mathcal{S}_{2}$ called channel state. Let $W_{Y|X_{1}X_{2}S_{1}S_{2}}(\cdot|\cdot)$ denote the channel transition probabilities.\footnote{$W_{Y|X_{1}X_{2}S_{1}S_{2}}(\cdot|\cdot)$ is abbreviated as $W_{Y|\underline{X}\underline{S}}(\cdot|\cdot)$} The evolution of $\underline{S}$ is iid across time with distribution  $W_{\underline{S}}$. Encoder $j$ is provided with the entire realization of component $S_{j}$ non-causally, and it's objective is to communicate message $M_{j}$ to the decoder. $M_{1},M_{2}$ are assumed to be independent and input $X_{j}$ is constrained to an average cost $\tau_{j}$ with respect to a cost function $\kappa_{j}: \mathcal{X}_{j}\rightarrow \mathbb{R}$.

The conventional technique for communicating over this channel - a MAC with channel state information distributed at transmitters (MAC-DSTx) - is to partition independent iid codes at each encoder and employ the technique of binning as is done for the PTP-STx channel in \cite{1980MMPCT_GelPin}. The decoder employs a joint typical decoder. Philosof and Zamir \cite{200906TIT_PhiZam} propose a new technique (PZ-technique) of correlated partitioning of coset codes for communicating over a binary \textit{additive} doubly dirty MAC-DSTx and prove that it strictly outperforms the conventional technique. In \cite{201301arXivMACDSTx_PadPra}, we generalized PZ-technique via union coset codes and derived a new achievable rate region $\beta_{f}(\ulinecost)$ for the general MAC-DSTx. In here, we provide a non-additive MAC-DSTx for which $\beta_{f}(\ulinecost)$ is strictly larger than $\alpha(\ulinecost)$, the largest known achievable rate region using iid codes.

\begin{figure}
\includegraphics[width=3.4in]{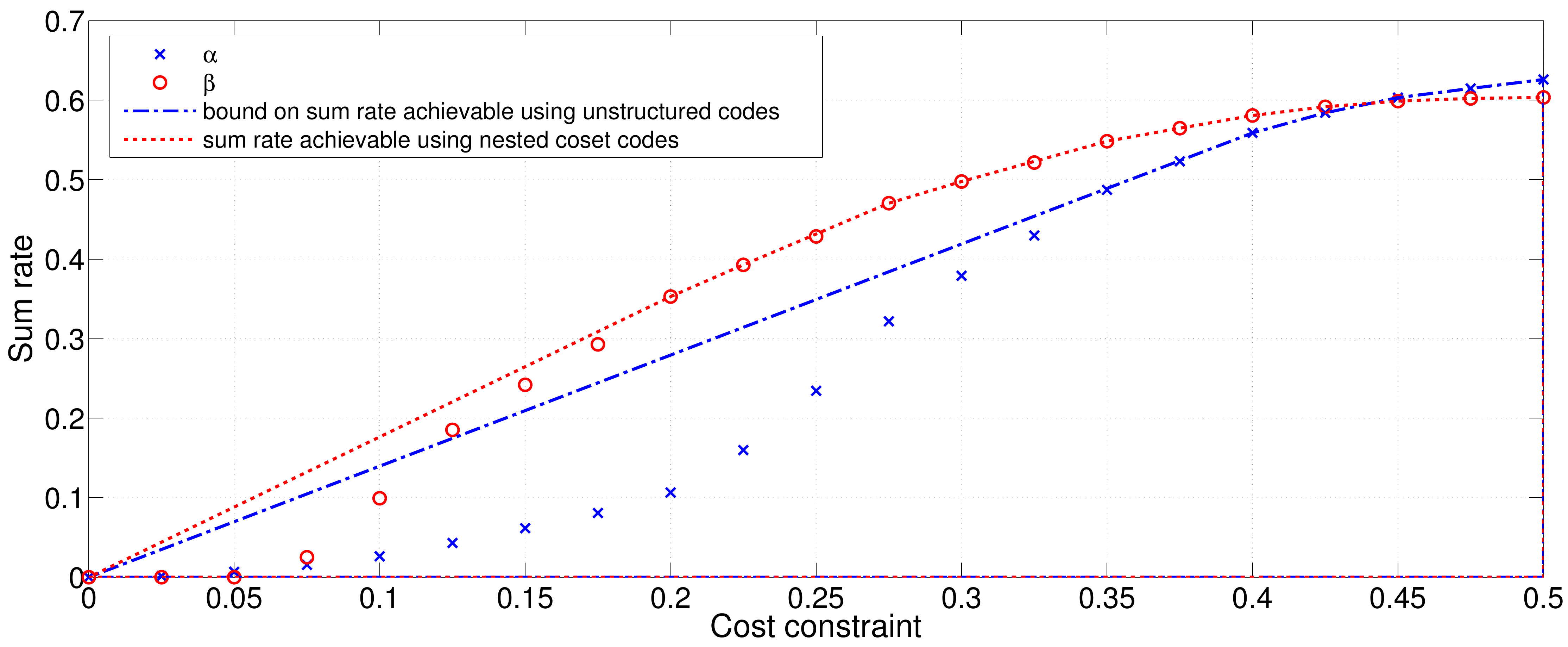}
\caption{Bounds on sum rate for example \ref{Ex:MAC-DSTx}}
\label{Fig:SumRatePlotsForIntractableChannel}
\end{figure}

\begin{example}
\label{Ex:MAC-DSTx}
Consider a binary MAC-DSTx with alphabet sets $\setS_{j}=\setX_{j}=\mathcal{Y}=\{ 0,1\}$, $j=1,2$, (ii)
uniform
and independent states, i.e.,
$W_{\underline{S}}(\underline{s}) = \frac{1}{4}$ for all $\underline{s} \in \underline{\mathcal{S}}$, (iii) and
Hamming cost function $\kappa_{j}(1,s_{j})=1$ and
$\kappa_{j}(0,s_{j})=0$ for any $s_{j} \in \setS_{j}$, $j=1,2$. The channel transition matrix is given in table
\ref{Table:ChannelTransitionMatrixExample5}. 1) An upper bound on sum rate achievable
using iid codes and 2) sum rate achievable using nested coset codes are
plotted in figure \ref{Fig:SumRatePlotsForIntractableChannel}.
\end{example}
\begin{table} \begin{center}
\begin{tabular}{|c|c|c|c|c|} \hline
$\scriptstyle X_{1}X_{2}S_{1}S_{2}$& $\scriptscriptstyle W_{ Y| \underline{XS}}(0|\cdot)$&
$\scriptstyle X_{1}X_{2}S_{1}S_{2}$ &$\scriptscriptstyle W_{ Y| \underline{XS}}(0|\cdot)$\\
\hline\hline
0000 & 0.92 & 0001 & 0.07  \\
\hline
1000 & 0.08 & 1001 &  0.92 \\
\hline
0010 & 0.06 & 0011 & 0.96 \\
\hline
1010 & 0.94 & 1011 &  0.10\\
\hline
0100 & 0.10 & 0101 &  0.88  \\
\hline
1100 & 0.92 & 1101 &   0.08\\
\hline
0110 & 0.95 & 0111 &  0.11\\
\hline
1110 & 0.06 & 1111 &  0.91\\
\hline
\end{tabular} \end{center}
\caption{Channel transition matrix Example \ref{Ex:MAC-DSTx}}
\label{Table:ChannelTransitionMatrixExample5} 
\end{table} 
\bibliographystyle{IEEEtran}
{
\bibliography{IEEEabrv,wisl}
}
\end{document}